\documentclass{article}

\usepackage{arxiv}

\usepackage[utf8]{inputenc} 
\usepackage[T1]{fontenc}    
\usepackage{hyperref}       
\usepackage{url}            
\usepackage{booktabs}       
\usepackage{amsfonts}       
\usepackage{nicefrac}       
\usepackage{microtype}      
\usepackage{lipsum}		
\usepackage{graphicx}
\usepackage{natbib}
\usepackage{doi}

\usepackage{tabularx}
\usepackage{booktabs}
\usepackage{array}
\usepackage{multirow}
\usepackage{enumitem}
\usepackage{xcolor}
\usepackage{tcolorbox} 
\tcbuselibrary{skins, breakable}
\tcbset{
  width=\textwidth,
  halign=justify,
  center,
  breakable,
  colbacktitle=gray,
  coltitle=white,
  fonttitle=\sffamily\small,
  colback=white,
  sharp corners, 
  size=title,
  boxrule=0.1mm,
  fontupper=\small,
}

\newlist{RQ}{enumerate}{1}
\setlist[RQ]{label=\textbf{RQ\arabic*:},ref={RQ\arabic*},itemsep=0ex,topsep=1pt}
\newcommand{\rqone}{How does accessing materials via RAG, as compared to direct access, influence learners' mode of following a plan?} 
\newcommand{\rqtwo}{How does accessing materials via RAG, as compared to direct access, influence learners' mode of monitoring?}
\newcommand{\rqthree}{How does accessing materials via RAG, as compared to direct access, influence learners' mode of exploring?}

\title{Changing the Optics: Comparing Traditional and Retrieval-Augmented GenAI E-Tutorials in Interdisciplinary Learning}

\author{
    {Hannah Kim}\\
	Department of Biology\\
	Temple University\\
	Philadelphia, PA 19122 \\
	\texttt{hannah.kim0007@temple.edu} \\
	\And
    {Rahad Arman Nabid} \\
	Dept. of Computer \& Information Sciences\\
	Temple University\\
	Philadelphia, PA 19122 \\
	\texttt{rahad.arman.nabid@temple.edu} \\
    \And
    {Jeni Sorathiya} \\
	Dept. of Computer \& Information Sciences\\
	Temple University\\
	Philadelphia, PA 19122 \\
	\texttt{jeni.sorathiya@temple.edu} \\
    \And
    {Minh Doan} \\
	School Of Arts And Sciences\\
	Rutgers University–New Brunswick\\
	New Brunswick, NJ 08901 \\
	\texttt{minh.n.doan@rutgers.edu} \\
    \And
    {Elijah Jordan} \\
	Dept. of Computer \& Information Sciences\\
	Temple University\\
	Philadelphia, PA 19122 \\
	\texttt{elijah.jordan@temple.edu} \\
    \And
    {Rayhana Nasimova} \\
	Dept. of Computer \& Information Sciences\\
	Temple University\\
	Philadelphia, PA 19122 \\
	\texttt{rayhana.nasimova@temple.edu} \\
    \And
    {Sergei L. Kosakovsky Pond} \\
	Department of Biology\\
	Temple University\\
	Philadelphia, PA 19122 \\
	\texttt{spond@temple.edu} \\
    \And
    {Stephen MacNeil} \\
	Dept. of Computer \& Information Sciences\\
	Temple University\\
	Philadelphia, PA 19122 \\
	\texttt{stephen.macneil@temple.edu} \\
}

\date{January 22, 2026}

\renewcommand{\headeright}{}
\renewcommand{\undertitle}{}
\renewcommand{\shorttitle}{Changing the Optics: Comparing Traditional and Retrieval-Augmented GenAI E-Tutorials in Interdisciplinary Learning}

\hypersetup{
pdftitle={Changing the Optics: Comparing Traditional and Retrieval-Augmented GenAI E-Tutorials in Interdisciplinary Learning},
pdfsubject={human-computer interaction, bioinformatics},
pdfauthor={Hannah Kim, Rahad Arman Nabid, Jeni Sorathiya, Minh Doan, Elijah Jordan, Rayhana Nasimova, Sergei L. Kosakovsky Pond, Stephen MacNeil},
pdfkeywords={generative AI, information search, interdisciplinary learning, informal learning},
}

\begin{document}
\maketitle

\begin{abstract}
	Understanding information-seeking behaviors in e-learning is critical, as learners must often make sense of complex and fragmented information, a challenge compounded in interdisciplinary fields with diverse prior knowledge. Compared to traditional e-tutorials, GenAI e-tutorials offer new ways to navigate information spaces, yet how they shape learners’ information-seeking behaviors remains unclear. To address this gap, we characterized behavioral differences between traditional and GenAI-mediated e-tutorial learning using the three search modes of orienteering. We conducted a between-subject study in which learners engaged with either a traditional e-tutorial or a GenAI e-tutorial accessing the same underlying information content. We found that the traditional users maintained greater awareness and focus of the information space, whereas GenAI users exhibited more proactive and exploratory behaviors with lower cognitive load due to the querying-driven interaction. These findings offer guidance for designing tutorials in e-learning.
\end{abstract}

\keywords{human-computer interaction \and bioinformatics \and generative AI \and information search \and interdisciplinary learning}

\section{Introduction}
Three years after the introduction of OpenAI’s GPT-4 \cite{achiam2023gpt}, generative AI (GenAI) is increasingly embedded in everyday life across a wide range of roles. In educational context, GenAI is commonly positioned as a tutor, placing particular emphasis on the quality and organization of information it provides. To support reliable access to learning materials, retrieval-augmented generation (RAG) has emerged as a common approach for grounding generated responses in high quality sources. In this work, we examine interactions mediated by a RAG-based system to better understand how learners engage with learning materials.

As GenAI becomes increasingly prevalent, more learners use GenAI systems for learning, and a growing body of work has examined these interactions in the context of search systems \cite{yang2025search+,pasquarelli2025ai}. However, many learners engage with learning materials in the form of traditional instructor-authored, text-based e-tutorials. For instance, community-driven databases of e-tutorials exist for tools and workflows in bioinformatics \cite{gruning2017jupyter, batut2018community, serrano2021fostering, gentleman2004bioconductor}. Users of bioinformatics tools are typically familiar with traditional e-tutorials as a primary means of obtaining information. Given the interdisciplinary nature of bioinformatics and the wide variation in learners’ prior knowledge, it is important to examine how the versatility of GenAI systems may provide additional support to learners accustomed to traditional e-tutorials.

In this study, we investigate how accessing e-tutorials via RAG changes learners' information-seeking behaviors through the conceptual framework of orienteering \cite{o1993orienteering}.
\begin{tcolorbox}[title=RESEARCH QUESTIONS]
\begin{RQ}
\setlist[RQ]{leftmargin=0pt}
    \item \rqone
    \item \rqtwo
    \item \rqthree
\end{RQ}
\end{tcolorbox}

Our contributions are empirical, artifact-based, and opinion-based \cite{wobbrock2016research}. 

\begin{tcolorbox}[title=CONTRIBUTIONS]
\begin{enumerate}[leftmargin=15pt]
    \item We captured differences in information-seeking behavior between users of the traditional and GenAI e-tutorials.
    \item We developed two types of e-tutorials for computer science learners about HyPhy \cite{pond2005hyphy}, a widely used bioinformatics tool with over 3,000 citations, and established a comparative evaluation framework.
    \item We present affordances of traditional and GenAI e-tutorials as potential considerations for designing e-tutorials.
\end{enumerate}
\end{tcolorbox}

\begin{figure*}[h] 
    \centering
    \includegraphics[width=\textwidth]{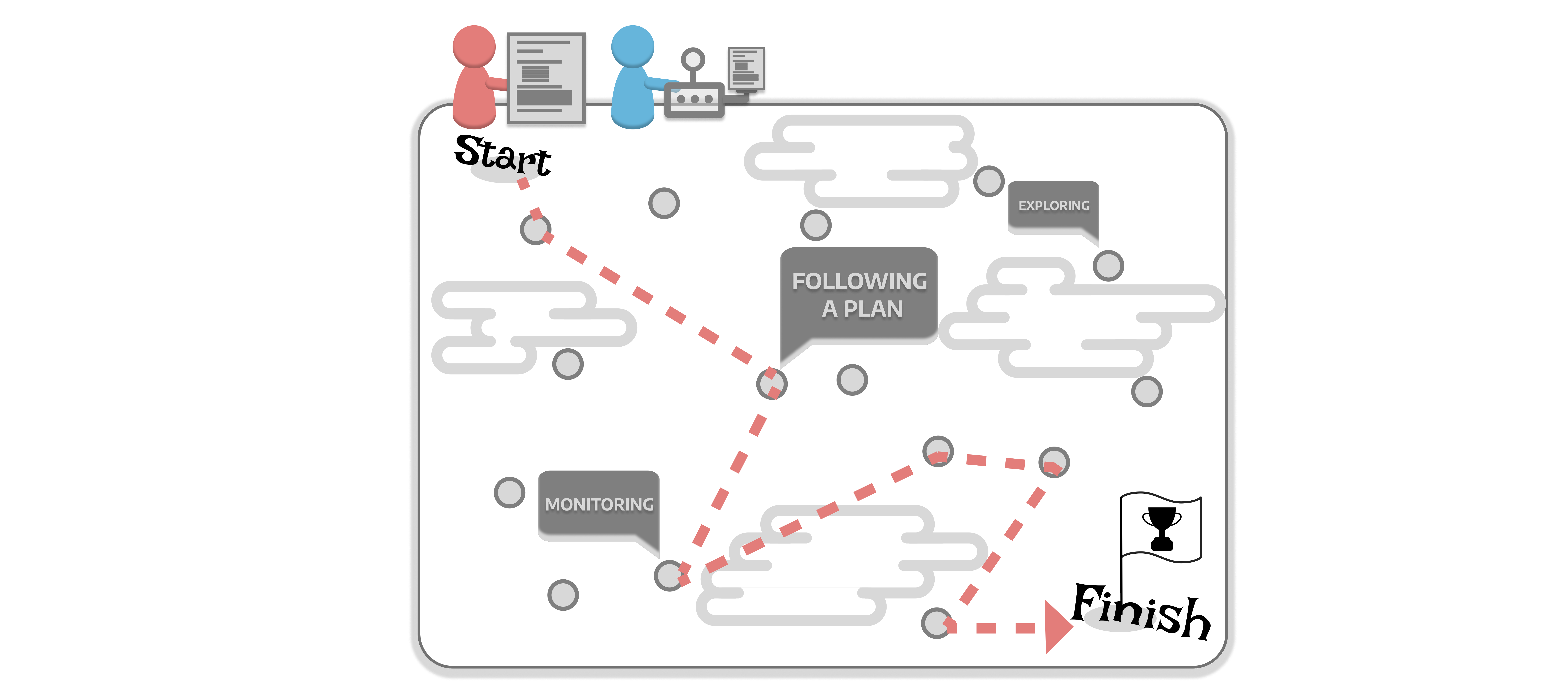}
    \caption{Conceptual Depiction of Orienteering \cite{o1993orienteering}. Users transition between states toward the finish line in the question-answering task by relying on either a traditional or a retrieval-augmented GenAI e-tutorial. Both tutorials provide access to the same information space, but differences in how the information is accessed distinguish them. This process of seeking information is analogous to navigation in a forest or fog, as users proceed without full knowledge of the underlying rules or state space.}
    \label{fig:1}
\end{figure*}

\section{Related Works}

Information-seeking behavior with GenAI has been widely studied. Much of the recent literature on information search adopts traditional search mediated by algorithmic information retrieval systems as the baseline. In educational contexts, researchers have compared student performance and perceptions when using traditional search versus GenAI systems, with some studies drawing on information from the World Wide Web \cite{yang2025search+} and others relying on course materials as the underlying information source \cite{pasquarelli2025ai}. Another line of work examines information search by focusing on the non-linear ways users navigate information landscapes, such as through the concept of orienteering. Although the original orienteering framework described information search mediated by human intermediaries \cite{o1993orienteering}, more recent work has extended this perspective to study search patterns mediated by web search engines and GenAI systems \cite{yen2025tosearch}.
 
\section{Methods}

\begin{figure*}[h] 
    \centering
    \includegraphics[width=\columnwidth]{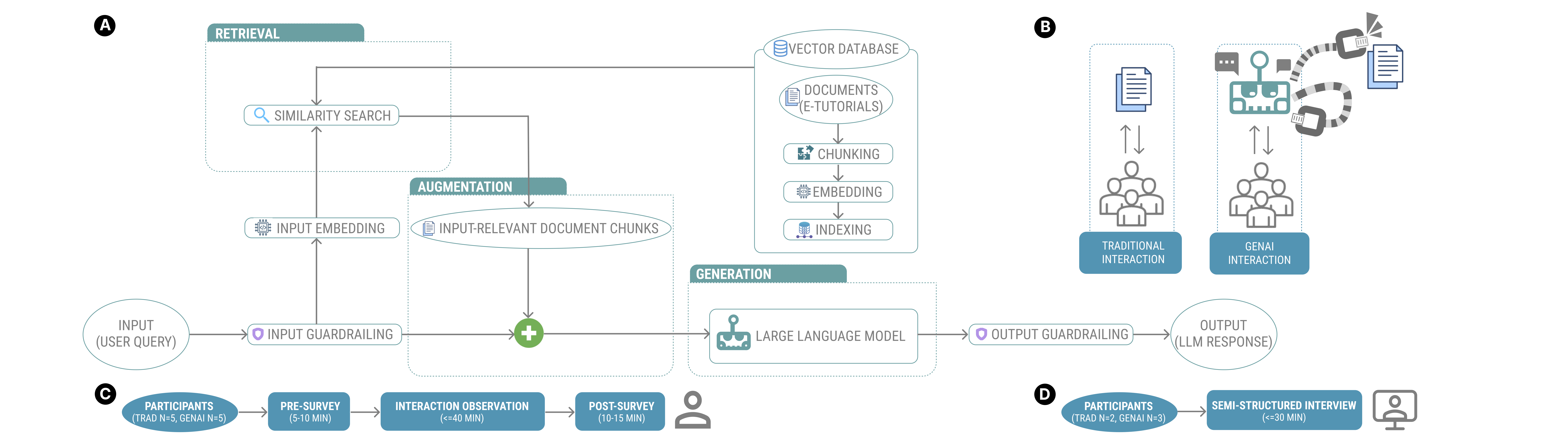}
    \caption{Study Overview: \textbf{(A)} Retrieval-Augmented Generation system diagram; \textbf{(B)} comparison of information-seeking behavior between a control group using a traditional e-tutorial and a treatment group using a retrieval-augmented large language model (hereafter GenAI) e-tutorial; \textbf{(C)} in-person survey and observation workflow; and \textbf{(D)} online interview workflow.}
    \label{fig:2}
\end{figure*}

To investigate how accessing information through GenAI influences users' orienteering behaviors, we conducted a between-subjects experimental study with 10 participants who are at least 18 years old, speak and read fluent English, studying computer science at Temple University, and have not previously used HyPhy \cite{kosakovsky2020hyphy,pond2005hyphy}. Data collection was limited to October 13--November 18, 2025 to ensure consistent participant perceptions of GenAI. The participants were recruited by posting advertisements on physical bulletin boards at Temple University and by asking instructors and organization leaders to disseminate the materials within their groups. Upon expressing interest, each participant reserved a time slot with the observer. Participants were compensated for their time at \$10 per hour rate in gift card. In addition, we implemented measures to minimize psychological risk from inappropriate GenAI responses and maximize privacy. The study protocol was approved as ethical and legal by Temple University Institutional Review Board.

\subsubsection{Design Principles} 
We designed our study considering information equivalence \cite{simon1978forms} and ecological relevance \cite{kieffer2017ecoval,brunswik1952conceptual,hammond2001essential} across the control and treatment conditions. Participants in the control group (N=5) interacted with a traditional e-tutorial, while those in the treatment group (N=5) interacted with a GenAI e-tutorial (See Figure \ref{fig:2}. Study Overview). Additional information on participant demographics, information-seeking behaviors \cite{o1993orienteering,pirolli1999information,kim2025conversations}, and related factors \cite{kim2025conversations,long2020ai,hart2006nasa,measuringu_nasa_tlx,nasa_tlx_scale,yang2025search+} was collected via surveys and interviews (See Appendix A).

\subsubsection{Traditional E-Tutorial} 
We developed a new traditional e-tutorial for HyPhy \cite{tradetutorialHK} to ensure that no state-of-the-art large language models had been trained on the content. Briefly, the tutorial describes where the software fits within the broader bioinformatics workflow, how to install it, how to run the methods, and how to interpret the outputs. The tutorial follows a conventional structure of an introduction, body, and conclusion. In addition, it was developed using metaphors and examples designed to appeal to students in computer science. The tutorial’s correctness and simplicity were confirmed by a senior researcher in bioinformatics, while its general readability was confirmed by a researcher without a background in biology or bioinformatics. The tutorial was provided on the HackMD platform \cite{hackmd} and was designed to be read in 10 to 30 minutes.

\subsubsection{GenAI E-Tutorial} 
The GenAI e-tutorial provided access to the traditional e-tutorial material by GenAI (See Figures \ref{fig:2}A\&\ref{fig:2}B). Because integration of retrieved information into the input layer is a simple and common way to provide additional context to the generator \cite{fan2024survey}, retrieved information, the user query, and the user chat history were combined and provided as input to the large language model in our implementation \cite{RA-BSTS_repo}. We used the Groq client \cite{gwennap2020groq} for the generator (\verb|llama-3.3-70b-versatile|, with knowledge cutoff in December 2023) \cite{meta_llama3.3_modelcard} and the guardrail \cite{huyen2025ai} (\verb|meta-llama/llama-guard-4-12b|) \cite{meta_llama_guard4_12b} models, and a local Ollama model \cite{ollama_website} for embedding (\verb|nomic-embed-text|) \cite{ollama_nomic_embed_text}. Chroma \cite{chroma_core_chroma} was used as the vector database with the LangChain framework \cite{langchain}. The guardrail and embedding models were selected based on popularity. The generator model was selected to minimize model bias on information related to HyPhy (See Figure \ref{fig:S1}. Model Selection for the Generator).

\subsubsection{Study Protocol} 
The study was conducted in two phases: an in-person phase (Figure \ref{fig:2}C) and an online follow-up interview phase (Figure \ref{fig:2}D). Data were collected from participants individually in both phases. In the first phase (See Figure \ref{fig:S2}. In-Person Experimental Setup), participants were randomly assigned to either the control or treatment group upon arrival at a private room. Participants were provided with a personal laptop to complete all activities during the in-person phase. After signing the informed consent form, participants completed a pre-survey. They then watched a video situating the task scenario in an ecologically relevant context (See Figure \ref{fig:S3}. Video Version of the Task Scenario) and interacted with the assigned e-tutorial while referring to a copy of the scenario for instructions and questions (See Figure \ref{fig:S4}. Written Version of the Task Scenario). Participants were encouraged to take notes and/or highlight contents \cite{palani2021conotate,yen2025tosearch,yang2025search+}, and were permitted to use external resources \cite{ahmad2015technology, yen2025tosearch}, as long as the tutorial remained their primary source of information. To streamline this process without information leakage among participants, the Chrome browser was configured to disable AI Overviews by default \cite{turnoffAIOverview}, and browsing history was cleared after each use. After completing the task, participants filled out a post-survey. Participant interaction with the e-tutorial was monitored via screen sharing while the observer remained outside the room. The observer intervened only when participants had questions, faced technical issues or attempted to run HyPhy on the computer. Screen recordings (without audio) and chat logs were collected during the observation. In the second phase, a semi-structured interview was conducted with a subset of the participants. Audio recordings of the interviews were collected.

\subsubsection{Data Analysis} 
We de-identified and analyzed the data using a mixed-methods approach. Quantitative and inductive qualitative methods were applied to the survey data to identify trends in participant impressions. Directed qualitative content analysis \cite{hsieh2005three,delve2020directed,kibiswa2019dqlca} was conducted on the observation and interview data to examine how our findings align with the concepts \cite{o1993orienteering,pirolli1999information} observed in prior work \cite{kim2025conversations}. Materials were annotated with predefined codes, which were made traceable through time stamps and analytic memos \cite{birks2008memoing}.

\section{Results}

Our findings reveal how GenAI-mediated e-tutorials influence learners’ orienteering behaviors relative to traditional e-tutorials. Our participants were aged 18--34 and were balanced in terms of gender (See Panel A of Figure \ref{fig:S5}. Bar Graphs of Quantitative Results). Participants varied in their years of study in computer science and were mostly unfamiliar with the field of biology or bioinformatics (See Figure \ref{fig:S5}B). Orienteering in an unfamiliar topic exerted cognitive pressure on participants, with both the traditional and GenAI groups consistently noting gaps in background knowledge. One participant decided early on that investing time in domain knowledge unattainable within the given time frame \textit{“would not be helpful,”} and adjusted their strategy accordingly. Another participant described the amount of information as overwhelming because they did not have relevant prior knowledge. Yet another participant became acutely aware of their lack of topic knowledge during the experiment: while the context of the COVID-19 pandemic was familiar and they recognized the term DNA, the topic was otherwise unfamiliar. However, the cognitive load differed between the traditional and GenAI groups. Participants in the traditional group reported mean NASA Task Load Index (NASA-TLX) \cite{measuringu_nasa_tlx,nasa_tlx_scale,hart2006nasa} scores at least 20 points higher (out of 100) than those in the GenAI group (See Figure \ref{fig:S5}E\textit{2}). The two observed outliers resulted from prior biology coursework in one participant and context-length constraints in the other. Although participants in the GenAI group experienced lower cognitive load, they still reported finding the topic challenging. In the following paragraphs, we investigate how participants' modes of search were shaped by different e-tutorials (See Table \ref{tab:S2}. Codebook for Directed Qualitative Content Analysis of the Information-Seeking Framework).

\subsubsection{RQ1: Following a plan}
Because participants were given a task or an outline of the required outcome, following a plan was the dominant state in both the traditional and GenAI groups. Participants enacted this mode differently within each group. In the traditional group, one participant spent most of their time reading the e-tutorial sequentially before revisiting the task scenario, while others alternated between the e-tutorial and the task scenario. In the GenAI group, all participants alternated between the GenAI e-tutorial and the task scenario but differed in how they initiated interactions with the GenAI: two asked how to run the input data, one asked about input data specifics, and two asked about HyPhy itself. The between-group difference stemmed from whether participants applied the concept of method selection, the core of the task scenario problem statement, when writing their answers. Three participants in the traditional e-tutorial group submitted answers with method selection in mind, compared to only one in the GenAI e-tutorial group. The GenAI group’s chat history indicated that most participants repeatedly received generic responses covering the topic. Only one participant who consistently narrowed the scope of the topic and occasionally performed external \textit{"cross-checks"} received responses with the level of specificity required by the task. Based on the interviews, participants were aware that GenAI responses could be incorrect; however, they had no reason to actively doubt the responses (See Figure \ref{fig:S5}\textit{G}) when participants \textit{"couldn't tell for sure"} that the responses were incorrect. Their experience differed from interacting with the \textit{"standard, normal, traditional tutorial, the one that [participants] come across very often,"} which is intentionally designed to guide learners and repeat information as needed for emphasis.

\subsubsection{RQ2: Monitoring}
Operationally, the monitoring mode was identified when participants tracked information they were already aware of. In both groups, this involved maintaining a reservoir of information in the form of notes (hereafter also referred to as memos), even if only briefly. Memos provided a virtual space for interacting with information and organizing thoughts. Taking memos was viewed as \textit{"having a backstory of the answer"}. Participants frequently used memos to record procedural details and salient information during formative and review states. Memos written by participants in the GenAI group were more expository and first-person in tone.

\subsubsection{RQ3: Exploring}
Participants in the GenAI group explored beyond the given topic, while those in the traditional group did not. Two participants both ventured outside the immediate topic at two distinct time points: early in the activity, when they explored the folder structure surrounding the input file of interest, and later during answer review, when they sought information driven by curiosity. A third participant provided insight into why this behavior may have been specific to the GenAI group. Because the participant expected to obtain answers quickly from the GenAI e-tutorial, they initiated interactions with greater latitude, beginning by \textit{"building a basis"} rather than immediately attempting to solve the task. 

\section{Discussion}

We investigated shifts in orienteering behavior when participants navigated a predefined information space using either a traditional e-tutorial or a GenAI e-tutorial. The information space was defined by the task scenario and the traditional e-tutorial. All participants had access to the task scenario; however, participants in the GenAI group accessed the e-tutorial indirectly. Both participants and the generator had minimal prior exposure to the information space, and the traditional e-tutorial was designed to be appropriate for the target users. Bioinformatics served as a suitable domain because it is a niche field that commonly attracts learners with a computer science background. The concept of orienteering \cite{o1993orienteering} previously inspired Yen et al. to compare patterns of information flow between web search engines and a GenAI system when both were used \cite{yen2025tosearch}. While we observed similar patterns, our work focuses on ensuring information equivalence across conditions to better understand the implications of using GenAI in learning.

\subsubsection{RQ1: Following a plan} 
Because participants were given a task, the orienteering trajectories in both groups were shaped around a plan to complete it, albeit differently. Participants in the traditional group viewed the entire information space upfront, whereas those in the GenAI group accessed it in slices per query, mediated by the GenAI system. This difference likely contributed to the distinct cognitive loads experienced by each group (See Figure \ref{fig:S5}E\textit{2}). Another key difference was tutorial design. Participants in the traditional group interacted with an intentionally designed e-tutorial, while those in the GenAI group effectively designed their own learning experience through the GenAI system. Awareness of the information space and the tutorial’s design intent likely helped the traditional group stay focused on the core problem, while the GenAI group needed to make conscious efforts in narrowing the scope.

\subsubsection{RQ2: Monitoring} 
Memoing was the primary form of information tracking in our study, naturally involving reflexivity \cite{birks2008memoing}. In the GenAI group, memos more often contained procedural information in an expository, first-person tone. This monitoring behavior underscores the proactivity required in the GenAI group, as querying necessitates active input from the user, unlike reading.

\subsubsection{RQ3: Exploring} 
The lower cognitive load observed in the GenAI group (see Figure \ref{fig:S5}E\textit{2}) reflects the greater latitude participants experienced, which made them more open to exploration during the experiment. While exploration of knowledge beyond the predefined information space could be directly observed during the experiment, recognition of opportunities to explore foundational concepts was captured only during interviews. Notably, participants were aware of the problem’s difficulty, yet they still ventured beyond the perceived scope driven by curiosity.

\section{Limitations and Future Work}
There are several limitations to our study. First, our sample size of 10 is too small to draw statistically significant conclusions. Second, the voluntary nature of the recruitment process has contributed to selection bias \cite{catalogue_of_bias_selection_bias,statisticshowto_bias} of participants. Our sample consisted of self-selecting participants who were motivated to engage; consequently, the findings should be interpreted as reflecting best-case scenarios in education \cite{anna_karenina_principle_wikipedia}. Third, our experimental condition for the GenAI e-tutorial may become obsolete as technologies quickly improve. Nevertheless, our study advances understanding of a fundamental shift and can inform the design of larger-scale studies on information spaces.

\section{Conclusion}

We characterized shifts in information-seeking behaviors when using the GenAI e-tutorial through the conceptual framework of orienteering. We found that acute awareness of the information space is an affordance of the traditional e-tutorial, whereas topic exploration and proactive question formulation are affordances of the GenAI e-tutorial. These findings offer guidance for incorporating GenAI systems into interdisciplinary learning and, more broadly, shed light on learning processes with GenAI. 

\section*{Acknowledgments}
We would like to thank our participants for their voluntary participation in our study. Our work would not have been possible without them. We would like to acknowledge C. Zastudil, A. Wazzan, E. Thyrum and S. Gupta for research-related suggestions, and H. Patel and S. Pandey for project team support. We thank K. Patel, S. Biswas, C. Pine-Simon, and A. Pang for their help with participant recruitment. We are also grateful to the Spring 2025 Senior Capstone project team (I. T. Applebaum, K. Orlovskiy, K. C. Bunn, A. Goldmeer, T. K. Witmer, J. Truong, I. Kabir, and K. Q. Nguyen) for their technical support. We would also like to thank the bioinformatics community and the computer and information sciences community (special thanks to the Temple HCI Lab and HCI Reading Group leads, A. Lazaro and A. Rahman) for their support and engagement. In this work, ChatGPT and Gemini were used to supplement grammar, text, and software development, but the core of the work is original. \\ \\ \\ \\ \\ \\ \\ \\ \\ \\ \\

\bibliographystyle{unsrt} 
\bibliography{preprint}  






\appendix 

\section{APPENDIX: Survey and Interview Questions} 
\subsection{Eligibility Screening Survey}
\begin{enumerate}
    \item We are looking for participants who meet the following criteria. Do you meet all four criteria below? \underline{(Yes/No)}
    \begin{itemize}
        \item of at least 18 years old, 
        \item speak and read fluent English, 
        \item studying computer science at Temple University, 
        \item and have not previously used HyPhy (https://hyphy.org/)
        .
    \end{itemize}
\end{enumerate}

\subsection{Pre-Survey (5-10 min)}
\begin{enumerate}
    \item (Demo)\footnote{Categorization of survey questions asked to participants. This information was not provided to participants during the survey.} What is your age \underline{(18-24/25-34/35-44/45-54/55+)}?
    \item (Demo) What is your gender identity \underline{(Man/Woman/Non-binary)}?
    \item (CS) How many years have you studied computer science since graduating from high school \underline{($<=$1 yr/2 yrs/3 yrs/4 yrs/5$<=$ yrs)}?
    \item (Bio) How many biology courses have you taken so far since graduating from high school \underline{(0/1/2/3/4+)}?
    \item (Bioinfo) How familiar are you with bioinformatics \underline{(Not Familiar/Slightly Familiar/Moderately Familiar/Familiar/Very Familiar)}?
    \item (Expected workload) How difficult do you expect this activity to be \underline{(Extremely Not/Somewhat Not/Neutral/Somewhat/Extremely)}?
    \item (Pre) What are your initial impressions about the e-tutorial assigned \underline{(Strongly Disagree/Disagree/Neutral/Agree/Strongly Agree)}?
    \begin{itemize}
        \item I have a positive impression of this type of e-tutorial.
        \item I am interested in the topics related to HyPhy.
        \item I am generally able to run any software successfully using this type of e-tutorial.
    \end{itemize}
    \item (AI literacy) \cite{long2020ai} How many times per week do you use generative AI tools like ChatGPT or Co-pilot \underline{(0/1-2/3-4/5-6/7+)}? 
    \item (AI literacy) How do you rate your knowledge on using generative AI tools \underline{(Very Poor/Poor/Fair/Good/Very Good)}? 
    \item (AI literacy) How confident are you in spotting errors when generative AI tools make mistakes \\ \underline{(Extremely not confident/Somewhat not confident/Neutral/Somewhat confident/Extremely confident)}?
\end{enumerate}

\subsection{Post-Survey (10-15 min)}
\begin{enumerate}
    \item (NASA-TLX) \cite{hart2006nasa, measuringu_nasa_tlx, nasa_tlx_scale} How would you rate the workload \underline{(Very Low/Low/Neutral/High/Very High)}?
    \begin{itemize}
        \item Mental demand: How mentally demanding was the task?
        \item Physical demand: How physically demanding was the task?
        \item Temporal demand: How hurried or rushed was the pace of the task? 
    \end{itemize}
    \item (NASA-TLX) How would you rate your performance \underline{(Perfect/Good/Neutral/Bad/Failure)}? In other words, how successful were you in accomplishing what you were asked to do?
    \item (NASA-TLX) How would you rate the effort and frustration level \underline{(Very Low/Low/Neutral/High/Very High)}?
    \begin{itemize}
        \item Effort: How hard did you have to work to accomplish your level of performance? 
        \item Frustration level: How insecure, discouraged, irritated, stressed, and annoyed were you?
    \end{itemize}
    \item (Post) What are your impressions about the e-tutorial \underline{(Strongly Disagree/Disagree/Neutral/Agree/Strongly Agree)}?
    \begin{itemize}
        \item Because of this experience, I have a more positive impression of this type of e-tutorial.
        \item Because of this experience, I have grown more interest in the topics related to HyPhy.
        \item Given this experience, I am confident that I can use HyPhy in the proper context.
        \item The e-tutorial provided sufficient content.
        \item The structure of the e-tutorial content was useful.
    \end{itemize}    
    \item (Perception) \cite{yang2025search+} How did you perceive the e-tutorial content \underline{(Strongly Disagree/Disagree/Neutral/Agree/Strongly Agree)}? 
    \begin{itemize}
        \item Reliable (or consistent)
        \item Credible (or believable)
        \item Unbiased (or objective)
        \item Factual (or verifiable)
        \item Trustworthy (or ethically dependable)
        \item Accurate (or correct)
    \end{itemize}
    \item (Verification) \cite{kim2025conversations} How often did you verify whether the content of the e-tutorial was true \underline{(Never/Rarely/Sometimes/Often/Always)}? 
    \item (Orienteering+Information Scent)  \cite{o1993orienteering,pirolli1999information,kim2025conversations} During the activity, I ... \underline{(Strongly Disagree/Disagree/Neutral/Agree/Strongly Agree/NA)}?
    \begin{itemize}
        \item Developed a learning strategy.
        \item Explored topics to my satisfaction.
        \item Adjusted my strategy when necessary.
        \item Used the e-tutorial structure as guidance.
        \item Used my prior knowledge in CS to understand topics.
        \item Used my prior knowledge in biology to understand topics.
    \end{itemize}
    
    \item (Orienteering+Information Scent) How did the e-tutorial support you? \underline{Please explain briefly.}
    \item (Orienteering+Information Scent) How did the e-tutorial challenge you? \underline{Please explain briefly.}
\end{enumerate}

\subsection{Semi-Structured Interview Questions (30 min)}
\subsubsection{Orienteering}\footnote{Categorization of interview questions asked to participants. This information was not provided to participants during the interview.} \cite{kim2025conversations, o1993orienteering}
\begin{enumerate}
    \item Can you describe your overall experience with the e-tutorial?
    \item How did you go about finding the information you needed for the task?
    \item To what extent, if at all, did you feel distracted by the features of the e-tutorial?
\end{enumerate}

\subsubsection{Information Scent} \cite{kim2025conversations, pirolli1999information}
\begin{enumerate}
    \item What kinds of hints or cues within the e-tutorial did you find especially useful?
    \item What characteristics made these hints helpful or unhelpful?
    \item Did you feel the number of hints was sufficient, insufficient, or overwhelming?
\end{enumerate}

\subsubsection{AI Literacy} \cite{kim2025conversations, long2020ai} 
\begin{enumerate}
    \item How confident were you in the quality and reliability of the e-tutorial content? \\ \\ \\ \\ \\ \\ \\ \\
\end{enumerate}

\section{APPENDIX: Supplementary Figures and Tables}

\setcounter{figure}{0}
\renewcommand{\figurename}{Supplementary Figure}
\renewcommand{\thefigure}{S\arabic{figure}}

\setcounter{table}{0}
\renewcommand{\tablename}{Table}
\renewcommand{\thetable}{S\arabic{table}}

\begin{figure*}[h] 
    \centering
    \includegraphics[width=\columnwidth]{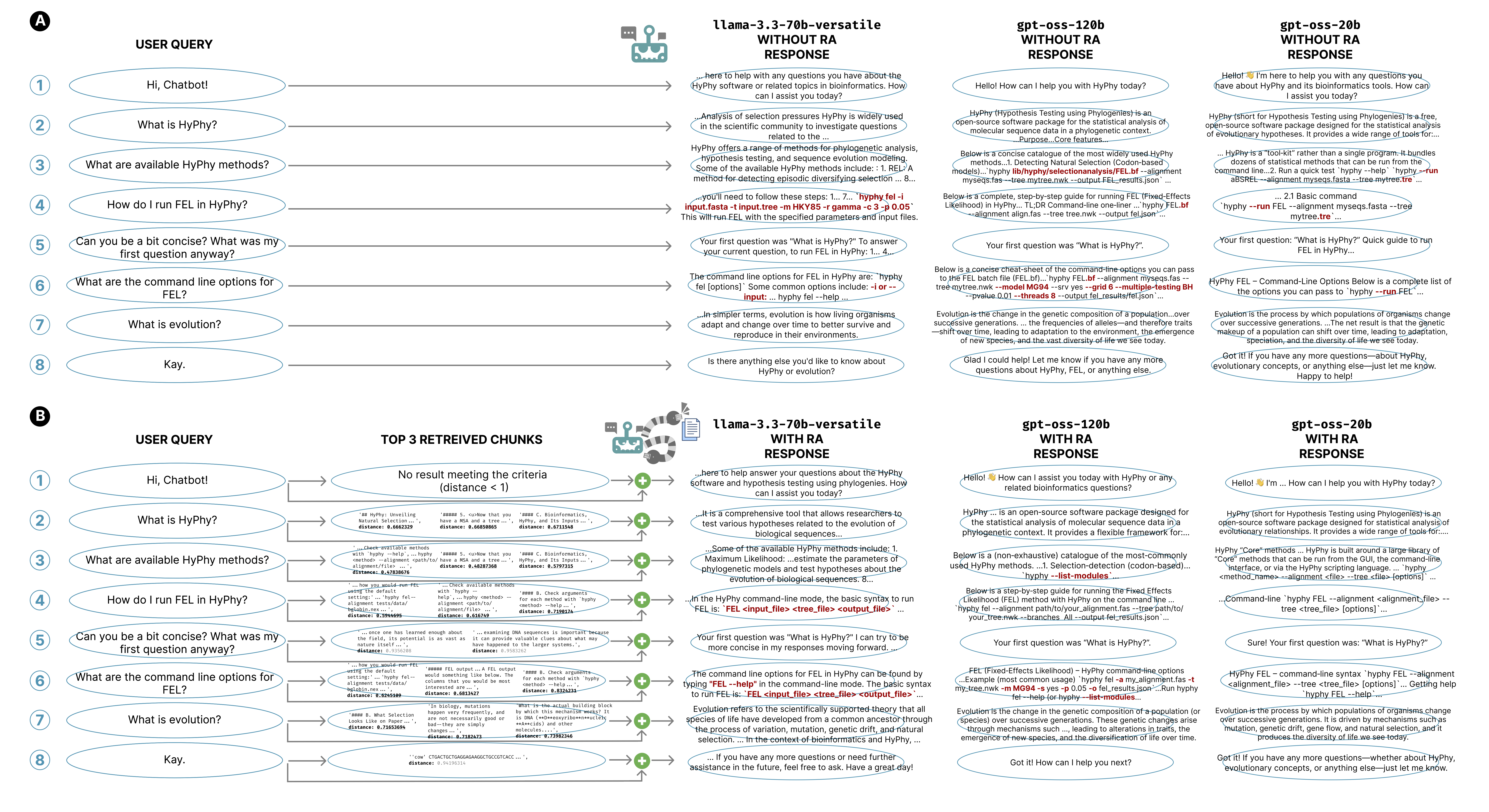}
    \caption{Model Selection for the Generator. To minimize potential model bias, we selected a model with limited prior domain-specific knowledge of HyPhy \cite{kosakovsky2020hyphy}. Task-relevant and task-irrelevant user queries were combined to evaluate three well-known large language models \cite{groq_models_docs, meta_llama3.3_modelcard, agarwal2025gpt}. Model performance was assessed both \textbf{(A)} without and \textbf{(B)} with retrieval augmentation. The top three retrieved chunks list shows partial document fragments returned based on their similarity to the user query. Because similarity is estimated using L2 (Euclidean) distance, a lower distance indicates higher relevance. The `nomic-embed-text` model \cite{ollama_nomic_embed_text} was used for embedding. The excerpts from the retrieved chunks and GenAI responses shown in the figure highlight limitations of casual input augmentation in GenAI. Incorrect information about HyPhy is highlighted in red. Based on the findings, the `llama-3.3-70b-versatile` model \cite{meta_llama3.3_modelcard} was selected, the data analysis plan was revised and the distance threshold was updated to 0.9 \cite{RA-BSTS_repo}.}
    \label{fig:S1}
\end{figure*}

\begin{figure*}[h] 
    \centering
    \includegraphics[width=\columnwidth]{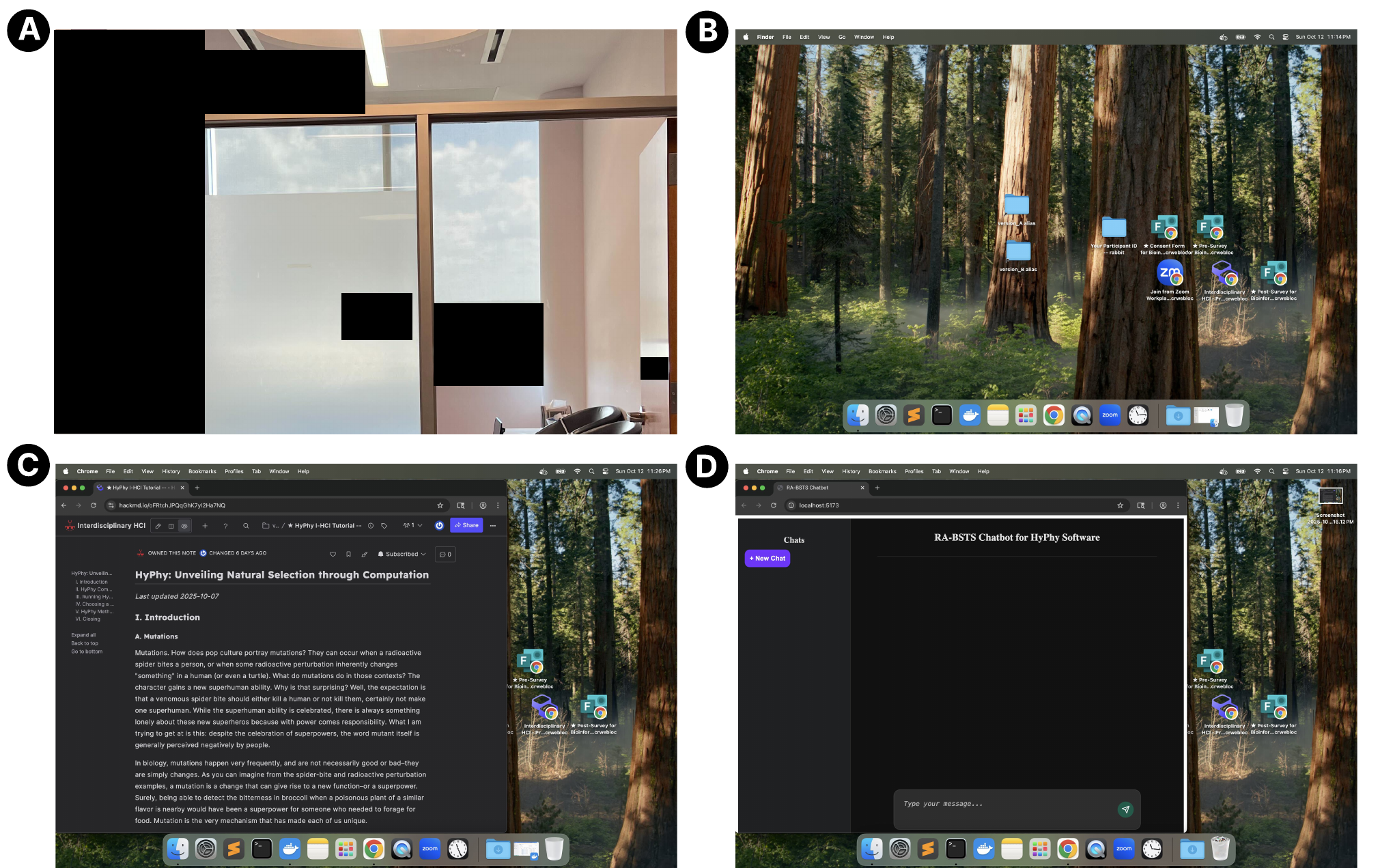}
    \caption{In-Person Experimental Setup: \textbf{(A)} a private room with a door and privacy glass that all in-person experiments were conducted; \textbf{(B)} the default computer screen; \textbf{(C)} the traditional e-tutorial screen; and \textbf{(D)} the GenAI e-tutorial screen.}
    \label{fig:S2}
\end{figure*}

\begin{figure*}[h] 
    \centering
    \includegraphics[width=\columnwidth]{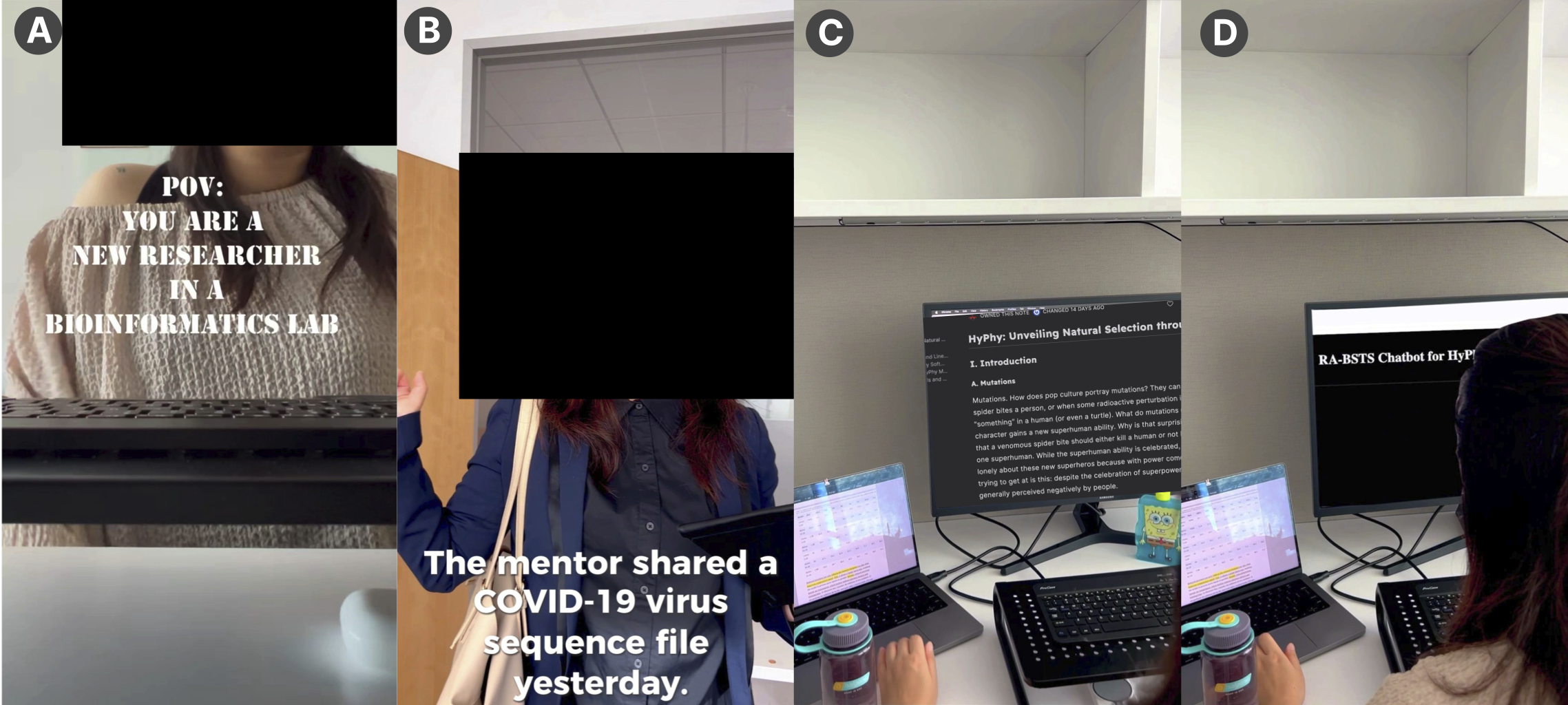}
    \caption{Video Version of the Task Scenario. An actor played dual roles of \textbf{(A)} a student researcher (the participant) and \textbf{(B)} a mentor in this point-of-view (POV) video. There were two versions of the video: \textbf{(C)} one that included the screen of the traditional e-tutorial, and \textbf{(D)} another that included the screen of the GenAI e-tutorial. Link to the video version of the task scenario \cite{youtube_shorts_Bnxgrbpujrk}: https://youtube.com/shorts/Bnxgrbpujrk.}
    \label{fig:S3}
\end{figure*}

\begin{figure*}[h] 
    \centering
    \includegraphics[width=\columnwidth]{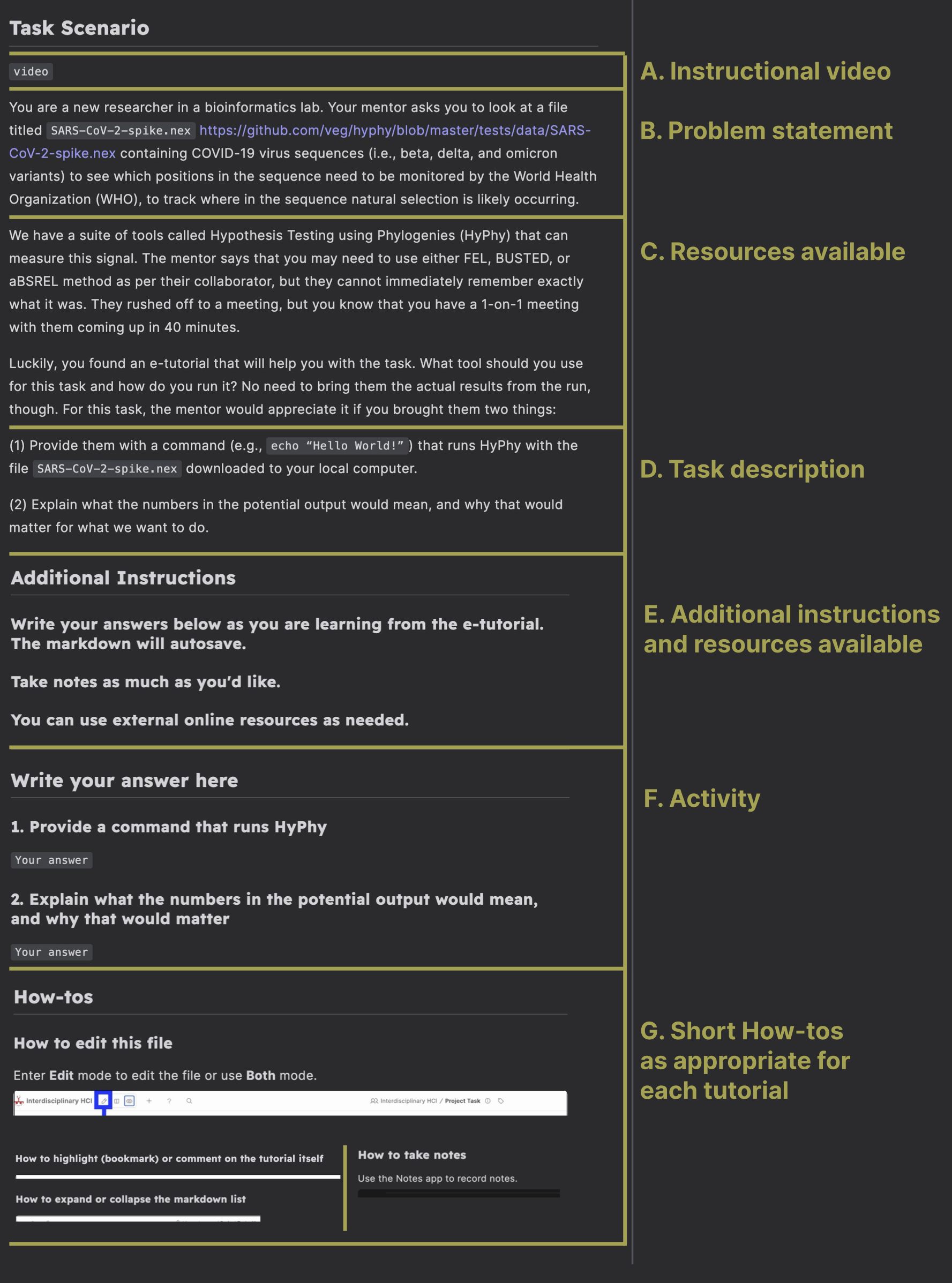}
    \caption{Written Version of the Task Scenario. It included instructions (A-G) that participants could refer to during their interactions.}
    \label{fig:S4}
\end{figure*}

\begin{figure*}[h] 
    \centering
    \includegraphics[width=\columnwidth]{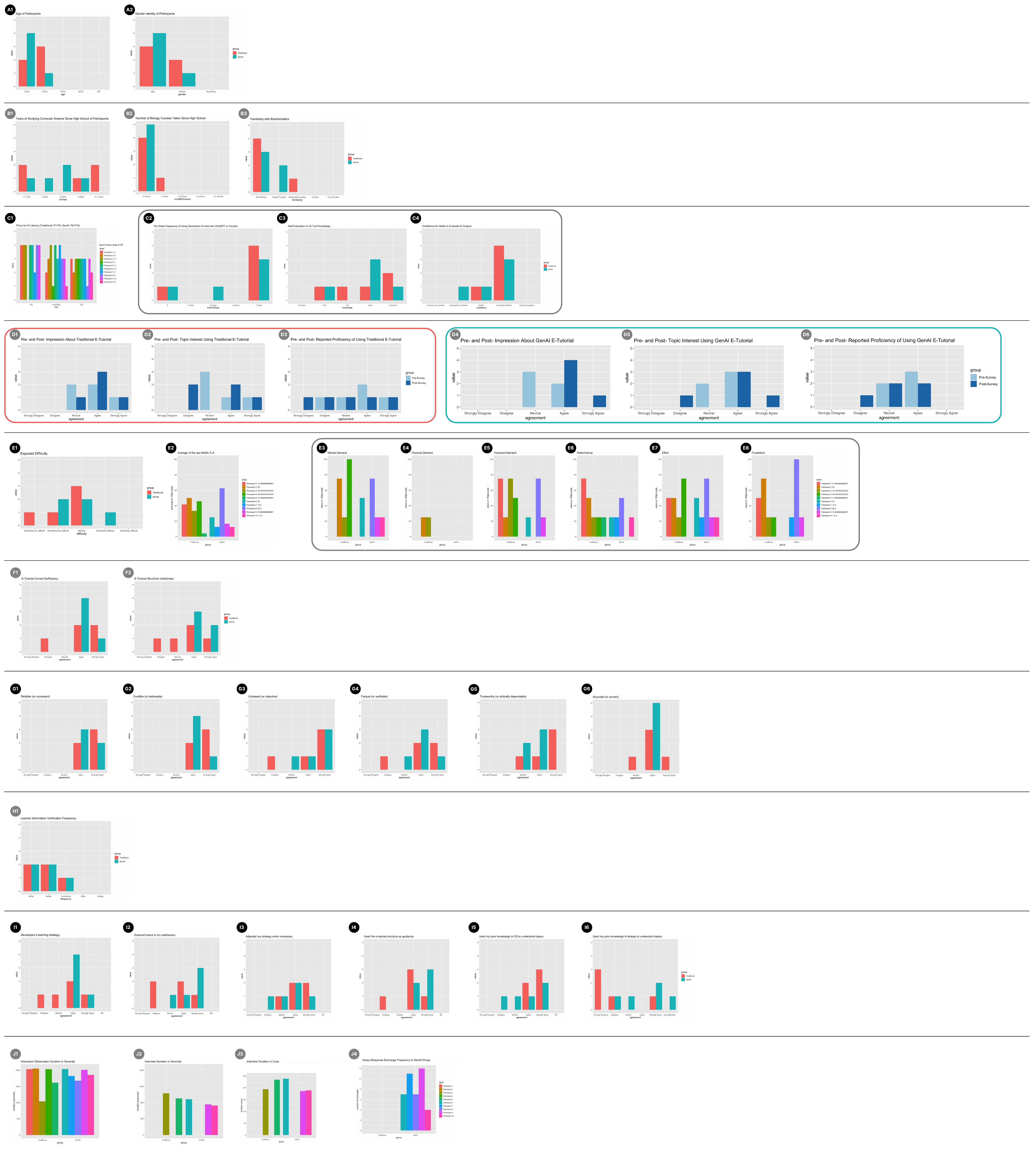}
    \caption{Bar Graphs of Quantitative Results: \textbf{(A)} demographics (pre-survey questions 1–2); \textbf{(B)} academic background (pre-survey questions 3–5); \textbf{(C)} AI literacy (pre-survey questions 8–10); \textbf{(D)} pre–post comparative perceptions (pre-survey question 7 and post-survey question 4); \textbf{(E)} cognitive loads (pre-survey question 6 and post-survey questions 1–3); \textbf{(F)} content sufficiency (post-survey question 4); \textbf{(G)} perceptions of content in six credibility categories (post-survey question 5); \textbf{(H)} verification frequency (post-survey question 6); \textbf{(I)} information-seeking behaviors (post-survey questions 7–9); and \textbf{(J)} time-related measurements taken during the study, including duration (in seconds and cues \cite{webvtt_wikipedia}) and query–response exchanges. Data were converted to 0--4 scale \cite{measuringu_nasa_tlx} for analysis when appropriate.}
    \label{fig:S5}
\end{figure*}

\begin{table*}[t]
\centering
\caption{Material List for Qualitative Analysis. Materials M1–M10 and M31–M35 served as the primary data sources for the directed qualitative content analysis. The mean screen-recording durations were 35:05 for the traditional group \textbf{(M1-M5)} and 37:37 for the GenAI group \textbf{(M6-M10)}. The GenAI group had a mean of 8.2 query–response exchanges with the e-tutorial \textbf{(M21-M25)}. Mean interview durations were 24:12 (214 cues in the audio transcription using WebVTT \cite{webvtt_wikipedia}) for the traditional group \textbf{(M31-M32)} and 19:43 (204 cues) for the GenAI group \textbf{(M32-M35)}. M12 is empty because one participant did not reach the question-answering stage.}
\label{tab:S1}
\tiny
\renewcommand{\arraystretch}{1.25}
\begin{tabularx}{\textwidth}{
>{\raggedright\arraybackslash}p{1.2cm}
>{\raggedright\arraybackslash}X
>{\raggedright\arraybackslash}p{2.4cm}
>{\raggedright\arraybackslash}p{2.2cm}
>{\raggedright\arraybackslash}p{3cm}
}
\toprule
\textbf{Material Number} &
\textbf{Materials for Qualitative Analysis} &
\textbf{Memos During Data Collection} &
\textbf{Media Type} &
\textbf{Size (Time / Exchanges / Time \& Cues)} \\
\midrule

M1 &
Screen recordings from Participant 1 &
Available &
Video &
40:42 \\

M2 &
Screen recordings from Participant 2 &
Available &
Video &
41:02 \\

M3 &
Screen recordings from Participant 3 &
Available &
Video &
20:48 \\

M4 &
Screen recordings from Participant 4 &
Available &
Video &
40:35 \\

M5 &
Screen recordings from Participant 5 &
Available &
Video &
32:19 \\

M6 &
Screen recordings from Participant 6 &
Available &
Video &
40:44 \\

M7 &
Screen recordings from Participant 7 &
Available &
Video &
36:27 \\

M8 &
Screen recordings from Participant 8 &
Available &
Video &
33:35 \\

M9 &
Screen recordings from Participant 9 &
Available &
Video &
40:16 \\

M10 &
Screen recordings from Participant 10 &
Available &
Video &
37:06 \\

M11 &
Task scenario answers and highlights from Participant 1 (reference for M1) &
N/A &
Short text &
N/A \\

M12 &
Task scenario answers and highlights from Participant 2 (reference for M2) &
N/A &
Short text &
N/A \\

M13 &
Task scenario answers and highlights from Participant 3 (reference for M3) &
N/A &
Short text &
N/A \\

M14 &
Task scenario answers and highlights from Participant 4 (reference for M4) &
N/A &
Short text &
N/A \\

M15 &
Task scenario answers and highlights from Participant 5 (reference for M5) &
N/A &
Short text &
N/A \\

M16 &
Task scenario answers and highlights from  Participant 6 (reference for M6) &
N/A &
Short text &
N/A \\

M17 &
Task scenario answers and highlights from Participant 7 (reference for M7) &
N/A &
Short text &
N/A \\

M18 &
Task scenario answers and highlights from Participant 8 (reference for M8) &
N/A &
Short text &
N/A \\

M19 &
Task scenario answers and highlights from Participant 9 (reference for M9) &
N/A &
Short text &
N/A \\

M20 &
Task scenario answers and highlights from Participant 10 (reference for M10) &
N/A &
Short text &
N/A \\

M21 &
Chat transcript (json) from Participant 6 (reference for M6) &
N/A &
Text with time stamps &
7 exchanges \\

M22 &
Chat transcript (json) from Participant 7 (reference for M7) &
N/A &
Text with time stamps &
11 exchanges \\

M23 &
Chat transcript (json) from Participant 8 (reference for M8) &
N/A &
Text with time stamps &
7 exchanges \\

M24 &
Chat transcript (json) from Participant 9 (reference for M9) &
N/A &
Text with time stamps &
12 exchanges \\

M25 &
Chat transcript (json) from Participant 10 (reference for M10) &
N/A &
Text with time stamps &
4 exchanges \\

M26 &
Server log including retrieved chunks from Participant 6 (reference for M6) &
N/A &
Text with time stamps &
N/A \\

M27 &
Server log including retrieved chunks from Participant 7 (reference for M7) &
N/A &
Text with time stamps &
N/A \\

M28 &
Server log including retrieved chunks from Participant 8 (reference for M8) &
N/A &
Text with time stamps &
N/A \\

M29 &
Server log including retrieved chunks from Participant 9 (reference for M9) &
N/A &
Text with time stamps &
N/A \\

M30 &
Server log including retrieved chunks from Participant 10 (reference for M10) &
N/A &
Text with time stamps &
N/A \\

M31 &
Transcribed and de-identified interview (vtt) from Participant 3 &
Available &
Text with cues and time stamps &
25:47 \& 194 cues \\

M32 &
Transcribed and de-identified interview (vtt) from Participant 5 &
Available &
Text with cues and time stamps &
22:38 \& 234 cues \\

M33 &
Transcribed and de-identified interview (vtt) from Participant 6 &
Available &
Text with cues and time stamps &
22:04 \& 238 cues \\

M34 &
Transcribed and de-identified interview (vtt) from Participant 9 &
Available &
Text with cues and time stamps &
18:55 \& 186 cues \\

M35 &
Transcribed and de-identified interview (vtt) from Participant 10 &
Available &
Text with cues and time stamps &
18:10 \& 189 cues \\

\bottomrule
\end{tabularx}
\end{table*}

\begin{table*}[t]
\centering
\caption{Codebook for Directed Qualitative Content Analysis of the Information-Seeking Framework. Themes of orienteering and information scent were broken down into codes, each accompanied by operational definitions and illustrative examples \cite{delve2023codebook, kibiswa2019dqlca}. Some examples overlap because certain chunks could be annotated with more than one code.}
\label{tab:S2}
\tiny
\renewcommand{\arraystretch}{1.3}
\begin{tabularx}{\textwidth}{
>{\raggedright\arraybackslash}p{2.5cm}
>{\raggedright\arraybackslash}p{1.8cm}
>{\raggedright\arraybackslash}p{3cm}
X
X
}
\toprule
\textbf{Theme} &
\textbf{Code Number} &
\textbf{Code Name} &
\textbf{Operational Definition} &
\textbf{Analytic Memo Examples} \\
\midrule

\multirow{5}{=}{A. Orienteering}
& C1 & Exploring
& Refers to undirected behaviors loosely related to the task scenario questions, likely displayed out of curiosity or amusement.
& Briefly skimmed <the input data>.nex file but immediately viewed more than five other files and folders in the github repo with more weight on the data. Other data files had shorter sample names and therefore a better display of the DNA sequences. \\

& C2 & Monitoring
& Refers to behaviors related to tracking identified information over time. 
& Switched back to the task scenario tab. Began taking notes to organize thoughts. Gathered clues currently available.  \\

& C3 & Following the default plan (Task Scenario)
& Refers to directed behaviors strictly related to answer the task scenario questions.
& Began from the top and read the e-tutorial displayed on the screen paragraph by paragraph, thoroughly absorbing the material as given. \\

& C4 & Distraction to orienteering efforts
& Refers to any distraction identified or experienced in orienteering.
& (1) The observer had to intervene to restart the GenAI tab, because GenAI was frozen from the excessive context length. 
(2) The amount of information in the GenAI response can be \textit{"daunting"} but having the ability to ask for clarification counterbalanced it. \\

& C5 & Selection of a method (Task Scenario Question 1)
& Refers to behaviors with intention towards solving the task scenario problem statement. 
& (1) Temporarily highlighted relevant keywords like \textit{"position"} on the task scenario.
(2) Wrote an answer to question 1 using the code snippet as a template. Understood the need to change the input file name for the answer, so updated it based on the task scenario instructions. \\

& C6 & Interpretation of the output (Task Scenario Question 2)
& Refers to behaviors with intention towards providing an interpretation to the output of running the selected method. 
& In the course of looking for information for question 1, found more relevant information for question 2 instead. Added the information about implication of having a small p-value. \\

& C7 & Utilization of internal resources for verification
& Refers to effort of fact-checking by referring to the given (traditional or GenAI) e-tutorial or self.
& (1) Scrolled to the answers, switched to the GenAI to check for the last time that the command was indeed what was being asked.
(2) Queried <answer to question 2> to ask if the answer displayed understanding of question 2. \\

& C8 & Utilization of external resources for verification 
& Refers to effort of fact-checking by referring to resources outside of the given (traditional or GenAI) e-tutorial or self. 
& (1) Opened a new tab to seek external resource on 'HyPhy method', made sure it existed.
(2) Began testing the command on the Terminal. The observer intervened to reorient the participant away from directly installing and running the software. \\

\midrule

\multirow{3}{=}{B. Information Scent}

& C9 & Utilization of structural or visual cues
& Refers to behaviors utilizing e-tutorial structure such as headers and examples following the concept description.
& (1) Used the within-tutorial hyperlink to a different section of the e-tutorial. This information did not add much value.
(2) The AI response was long, but the length was quickly gauged by scroll bounce effect. \\

& C10 & Utilization of prior-knowledge based cues
& Refers to behaviors finding a contextual cue potentially valuable because of the presence or absence of prior knowledge.
& (1) Began testing the command on the Terminal. The observer intervened to reorient the participant away from directly installing and running the software.
(2) Focused on the code snippet. Re-read the related text.
(3) The word \textit{"null hypothesis"} appeared 13 times in the response to query about p-value. Queried about null hypotheses. \\
\bottomrule
\end{tabularx}
\end{table*}

\end{document}